\documentclass[11pt]{article}

\usepackage{amsmath}
\usepackage{graphicx}
\usepackage{amsfonts}
\usepackage{amssymb}
\usepackage{epsfig}
\usepackage{color}
\usepackage{psfrag}
\usepackage{epstopdf}

\usepackage{caption}
\usepackage{subcaption}

\newcommand{\EQ}{\begin{equation}}
\newcommand{\EN}{\end{equation}}
\newcommand{\be}{\begin{equation}}
\newcommand{\ee}{\end{equation}}
\newcommand{\bea}{\begin{eqnarray}}
\newcommand{\eea}{\end{eqnarray}}

\begin{document} \setcounter{page}{0}
\newpage
\setcounter{page}{0}
\renewcommand{\thefootnote}{\arabic{footnote}}
\newpage
\begin{titlepage}
\begin{flushright}
\end{flushright}
\vspace{0.5cm}
\begin{center}
{\large {\bf Nonuniversality in random criticality}}\\
\vspace{1.8cm}
{\large Gesualdo Delfino}\\
\vspace{0.5cm}
{\em SISSA -- Via Bonomea 265, 34136 Trieste, Italy}\\
{\em INFN sezione di Trieste, 34100 Trieste, Italy}\\
\end{center}
\vspace{1.2cm}

\renewcommand{\thefootnote}{\arabic{footnote}}
\setcounter{footnote}{0}

\begin{abstract}
\noindent
We consider $N$ two-dimensional Ising models coupled in presence of quenched disorder and use scale invariant scattering theory to exactly show the presence of a line of renormalization group fixed points for any fixed value of $N$ other than 1. We show how this result relates to perturbative studies and sheds light on numerical simulations. We also observe that the limit $N\to 1$ may be of interest for the Ising spin glass, and point out potential relevance for nonuniversality in other contexts of random criticality.
\end{abstract}
\end{titlepage}

\newpage
\tableofcontents

\section{Introduction}
The renormalization group (RG) \cite{Wilson} founds our understanding of diffferent areas of modern physics. In particular, the observed universal critical behavior of classes of microscopically different statistical systems is explained by their RG flow towards a same fixed point at large distances. On the other hand, while universality classes associated to isolated fixed points cover the majority of critical phenomena, RG theory also accounts for exceptions. These arise when, for a fixed symmetry of the Hamiltonian, a continuous manifold of fixed points is present. If this happens, the system can renormalize to different points on the manifold depending on its parameters, thus exhibiting continously varying critical indices and, in this sense, nonuniversality. The conditions of ``true marginality" required for the presence of a manifold of RG fixed points are rarely fulfilled but, if they are, implications may be particularly relevant. The prominent example known so far is the line of fixed points allowed by the two-dimensional Gaussian model, which accounts for the Berezinskii-Kosterlitz-Thouless (BKT) transition \cite{Berezinskii,KT}, the Luttinger liquid behavior of electrons in ($1+1$) dimensions \cite{Tomonaga,Luttinger,ML}, the continuously varying exponents of the Ashkin-Teller model \cite{Baxter,KB}, and other phenomena. 

In this paper, we show that nonuniversality due to the presence of a line of RG fixed points occurs also in the context of random criticality. This finding had been prevented so far by the fact that the notorious theoretical difficulty to deal with quenched disorder adds to that of establishing true marginality. In recent years, however, it has been shown that the RG fixed points of two-dimensional disordered systems with short range interactions can be {\it exactly} determined in the scattering framework \cite{random,DT2,DL_ON1,DL_ON2,DL_softening} (\cite{colloquium} for a review). Here we perform this analysis for the system of $N$ two-dimensional Ising models coupled in presence of disorder and show that a line of fixed points is present for any fixed $N$ other than 1. We explain how this finding relates to perturbative studies \cite{DD_AT,Dotsenko_AT,Murthy,Cardy_GN} and sheds light on numerical results \cite{WD,Katzgraber1,Katzgraber2,Vojta_3color,Vojta_4color}. We also observe that the limit $N\to 1$ may be of interest for the Ising spin glass, and point out potential relevance for nonuniversality in other contexts of random criticality.

In the next section we introduce the model and derive its exact fixed point equations. In section~\ref{solutions} we give the solutions of these equations for the pure case and that yielding the line of random fixed points. RG flows are examined in section~\ref{RG_flows} before discussing the results in the final section.

\section{Exact fixed point equations}
We consider $N$ two-dimensional Ising models coupled via an Hamiltonian invariant under spin reversal for each model and permutations among the models; the system with these symmetries is considered in presence of quenched disorder. A basic lattice Hamiltonian with these properties is
\EQ
H=-\sum_{\langle x,y\rangle}\left[J_{xy}\sum_{a=1}^N\sigma_a(x)\sigma_a(y)+K\sum_{a\neq b}\sigma_a(x)\sigma_a(y)\sigma_b(x)\sigma_b(y)\right]\,,
\label{lattice}
\EN
where $\sigma_a(x)=\pm 1$ is the spin variable of the $a$-th Ising model at site $x$ of a regular lattice, $\sum_{\langle x,y\rangle}$ denotes the sum over nearest neighbors, and disorder is introduced through random couplings $J_{xy}$ drawn from some probability distribution $P(J_{xy})$. The Hamiltonian (\ref{lattice}) corresponds to the (disordered) Ashkin-Teller model \cite{AT} for $N=2$, and to what is often called the $N$-color Ashkin-Teller model \cite{GW} for $N$ generic. On the other hand, as will become immediately clear, our derivation of exact RG fixed points equations is performed directly in the continuum implementing the above mentioned symmetries, so it will also include the fixed points of (short range) lattice realizations of these symmetries other than (\ref{lattice}), in particular those with different realizations of disorder (e.g. $K$ also random or random site dilution).  

In the continuum, RG fixed points correspond to conformal field theories (CFTs). We then use the scattering framework of \cite{paraf}, in which one of the two spatial dimensions plays the role of imaginary time. The method exploits the fact that infinite-dimensional conformal symmetry of two-dimensional RG fixed points \cite{DfMS} yields infinitely many quantities to be conserved in scattering processes, so that initial and final states are forced to be kinematically identical (complete elasticity). In addition, scale and relativistic invariance lead to scattering amplitudes that do not depend on energy. These drastic simplifications make the scattering problem exactly solvable, a circumstance that allowed to progress with difficult problems of two-dimensional criticality, both with \cite{random,DT2,DL_ON1,DL_ON2,DL_softening,colloquium} and without disorder \cite{DT1,ising_vector,DDL_nematic,DLD_RPN,DLD_CPN,potts_qr,RPN_universality}.

\begin{figure}[t]
\centering
\includegraphics[width=14cm]{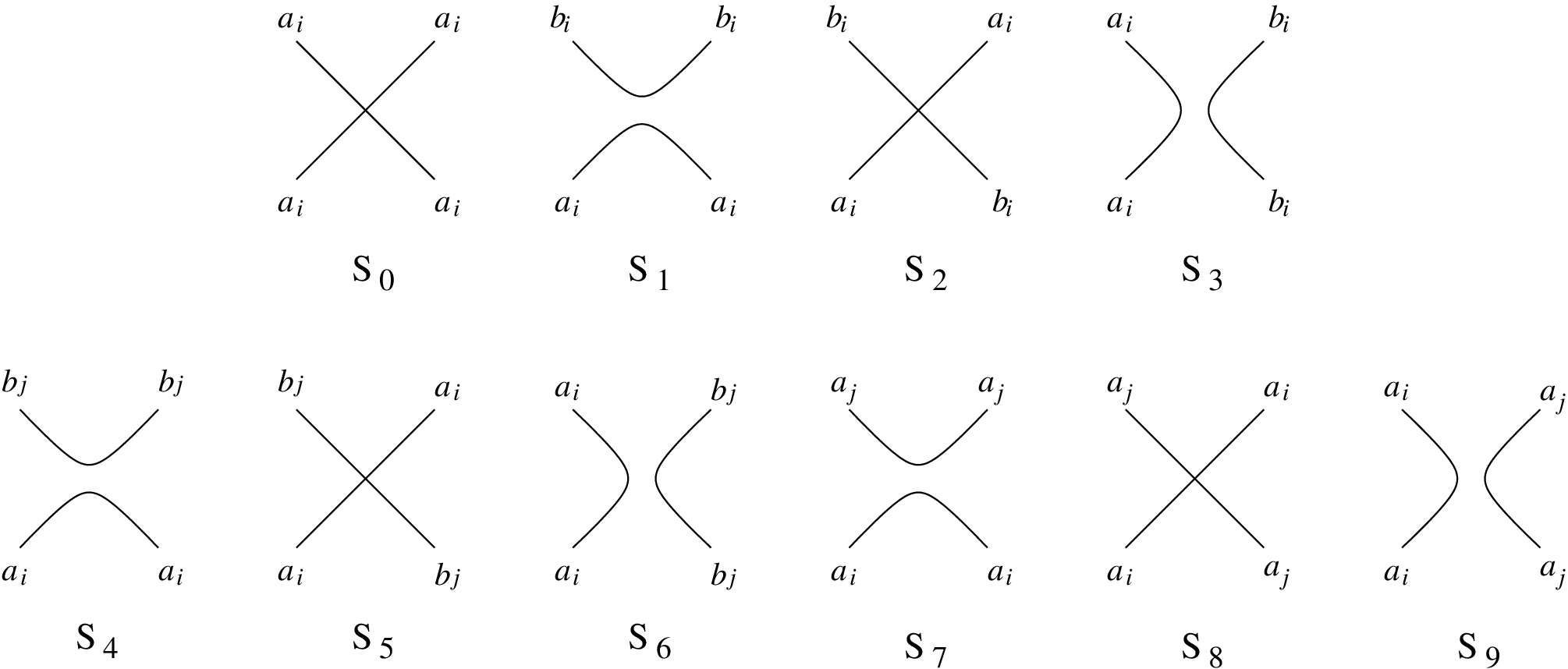}
\caption{Scattering amplitudes for the system of $N$ coupled Ising models with quenched disorder. $a_i$ labels a particle excitation in the $i$-th replica of the $a$-th Ising model ($a\neq b$, $i\neq j$). Time runs upwards.}
\label{amplitudes}
\end{figure}

The specification of the model follows from the symmetry representation carried by the scattering particles, which correspond to the fundamental collective excitation modes of the system. In our present case, we denote by $a=1,2,\ldots,N$ the particle corresponding to an elementary excitation in the $a$-th Ising model; this particle is odd under $\sigma_a\to-\sigma_a$. As usual (see e.g. \cite{Cardy_book}), quenched disorder is introduced considering $n$ replicas of the system, with $n$ to be eventually sent to zero. In the scattering framework this amounts to introducing a replica index $i=1,2,\ldots,n$, and labeling by $a_i$ a particle in the $i$-th replica of the $a$-th Ising model. It follows that the elastic scattering amplitudes allowed by spin reversal symmetry in each of the $Nn$ Ising models are those shown in Figure~\ref{amplitudes}. The amplitudes in the first row involve a single replica and are the only ones defined in the ``pure" model, namely the model without disorder ($n=1$). The amplitudes in the second row, instead, couple different replicas and enter the game in the disordered case. 

Adopting the notation $S_{\alpha\beta}^{\gamma\delta}={}_\alpha^\delta\times_\beta^\gamma$ for a generic amplitude, the energy-independence of the amplitudes leads to the particularly simple form of the crossing and unitarity equations \cite{paraf,colloquium}
\begin{equation}
S_{\alpha\beta}^{\gamma\delta}=[S_{\alpha\delta}^{\gamma\beta}]^*\,,
\label{cross}
\end{equation} 
\begin{equation}
\sum_{\epsilon,\phi} S_{\alpha\beta}^{\epsilon\phi}[S_{\epsilon\phi}^{\gamma\delta}]^*=\delta_{\alpha\gamma}\delta_{\beta\delta}\,,
\label{unitarity}
\end{equation}
respectively. The amplitudes are also invariant under spatial reflection ($S_{\alpha\beta}^{\gamma\delta}=S_{\beta\alpha}^{\delta\gamma}$) and time reversal ($S_{\alpha\beta}^{\gamma\delta}=S^{\alpha\beta}_{\gamma\delta}$).
For the amplitudes of Figure~\ref{amplitudes} the crossing equations (\ref{cross}) translate into
\begin{align}
S_k&=S_k^*\,,\hspace{3.4cm}k=0,2,5,8
\label{crossing1}\\
S_k&=S_{k+2}^*\equiv X_k+iY_k\,,\hspace{1cm}k=1,4,7\,,
\label{crossing2}
\end{align}
where we introduced $X_k$ and $Y_k$ real. Then the unitarity equations (\ref{unitarity}) take the form
\begin{align}
&S_0^2+(N-1)(X_1^2+Y_1^2)+(N-1)(n-1)(X_4^2+Y_4^2)+(n-1)(X_7^2+Y_7^2)=1\,,
\label{uni1}\\
&2S_0 X_1+(N-2)(X_1^2+Y_1^2)+2(n-1)(X_4 X_7+Y_4 Y_7)+(N-2)(n-1)(X_4^2+Y_4^2)=0\,,\\
&X_1 S_2=0\,,\label{uni3}\\
&X_1^2+Y_1^2+S_2^2=1\,,
\label{uni4}\\
&2 S_0 X_4+2(X_1X_7+Y_1Y_7)+2(N-2)(X_1X_4+Y_1Y_4)+2(n-2)(X_4X_7+Y_4Y_7)\nonumber\\
&+(N-2)(n-2)(X_4^2+Y_4^2)=0\,,\\
&X_4 S_5=0\,,\label{uni6}\\
&X_4^2+Y_4^2+S_5^2=1\,,\\
&2 S_0X_7+(n-2)(X_7^2+Y_7^2)+2(N-1)(X_1X_4+Y_1Y_4)+(N-1)(n-2)(X_4^2+Y_4^2)=0,\\
&X_7 S_8=0\,,\label{uni9}\\
&X_7^2+Y_7^2+S_8^2=1\,.
\label{uni10}
\end{align}

It follows from their derivation that the solutions of Eqs.~(\ref{uni1})-(\ref{uni10}) are the RG fixed points of the replicated system of $N$ Ising models with couplings constrained only by the requirement of preserving global invariance under spin reversals $\sigma_{a,i}\to -\sigma_{a,i}$ and permutations $\sigma_{a,i}\leftrightarrow\sigma_{b,j}$. Notice that the equations contain $N$ and $n$ as parameters that can take continuous values and that, in particular, the limit $n\to 0$ corresponding to quenched disorder can be taken straightforwardly.

It is also useful to notice that the theory we are considering reduces to the replicated $O(N)$ theory \cite{DL_ON1,DL_ON2,colloquium} when the scattering processes of Figure~\ref{amplitudes} do not distinguish $a\neq b$ from $a=b$. Hence, the $O(N)$-invariant case is obtained when
\EQ
O(N):\hspace{.5cm}S_0=S_1+S_2+S_3\,,\hspace{1cm}S_4=S_7\,,\hspace{1cm}S_5=S_8\,,\hspace{1cm}S_6=S_9\,.
\label{ON}
\EN

\section{Solutions}
\label{solutions}
\subsection{Pure case}
For $n=1$, Eqs.~(\ref{uni1})-(\ref{uni4}) receive contributions only from the single-replica amplitudes $S_0,\ldots,S_3$ and decouple from the remaining equations. Hence, (\ref{uni1})-(\ref{uni4}) with $n=1$ are the fixed point equations for the pure model. Their solutions are listed in Table~\ref{pure_solutions}. 

P1 with $S_2=S_0$, P2 and P3 are the fixed points of the $O(N)$-invariant subspace satisfying $S_0=S_1+S_2+S_3$ (recall (\ref{ON})). These solutions and their CFT characterization have been discussed in Refs.~\cite{DL_ON2,colloquium}, to which we refer the reader for the details; here we only recall some main points. P1$_{\pm}$ is the free boson/fermion solution which exists for any $N$. P2, defined for $N\in[-2,2]$, is the solution corresponding to the gas of self-avoiding loops \cite{Nienhuis}. P3 exists only for $N=2$ and is the line of fixed points parametrized by $Y_1$ responsible for the continuously varying exponents of the Ashkin-Teller model \cite{Baxter,KB} and for the BKT transition of the XY model \cite{Berezinskii,KT}. $Y1$ is related as $Y1=-\sin(\pi/x_\varepsilon)$ \cite{paraf} to the scaling dimension $x_\varepsilon$ of the energy density field $\varepsilon$ of the Ashkin-Teller model, which is also the most relevant $O(2)$-invariant field of the XY model; the BKT transition occurs when $\varepsilon$ becomes marginal ($x_\varepsilon=2$, i.e. $Y_1=-1$). 

For $N=2$ there are also two lines of fixed points, P4 and P5, not belonging to the $O(N)$-invariant subspace. The presence in the Ashkin-Teller model of three lines of fixed points having in common the point $Y_1=-1$ where $\varepsilon$ becomes marginal\footnote{The maximal value of $x_\varepsilon$ realized in the square lattice Ashkin-Teller model is 3/2 \cite{KB}.} was argued perturbatively in \cite{JKKN,Kadanoff} and already exactly determined through scale invariant scattering theory in \cite{ising_vector}.

\begin{table}
\begin{center}
\begin{tabular}{c|c|c|c|c|c}
\hline 
Solution & $N$ & $S_0$ & $S_2$ & $X_1$ & $Y_1$\\ 
\hline \hline
$\text{P}1_{\pm}$ & $(-\infty,\infty)$ & $\pm 1$ & $(\pm) 1$ & $0$ & $0$ \\ 
$\text{P}2_{\pm}$ & $[-2, 2]$ & $\pm\sqrt{2-N}$ & $0$ & $\pm\frac{1}{2}\sqrt{2-N}$ & $(\pm)\frac{1}{2}\sqrt{2+N}$ \\ 
$\text{P}3_{\pm}$ & $2$ & $\pm\sqrt{1-Y_1^2}$ & $\pm\sqrt{1-Y_1^2}$ & $0$ & $[-1,1]$ \\
$\text{P}4_{\pm}$ & $2$ & $\pm\sqrt{1-Y_1^2}$ & $\mp\sqrt{1-Y_1^2}$ & $0$ & $[-1,1]$ \\
$\text{P}5_{\pm}$ & $2$ & $0$ & $0$ & $\pm\sqrt{1-Y_1^2}$ & $[-1,1]$ \\
[0.7em] 
\hline 
\end{tabular} 
\caption{Solutions of Eqs.~(\ref{uni1})-(\ref{uni4}) with $n=1$. They correspond to the RG fixed points of the pure model. Signs in parenthesis are both allowed.
}
\label{pure_solutions}
\end{center}
\end{table}

\subsection{Disordered case}
The RG fixed points for the case of quenched disorder correspond to the solutions of the exact fixed point equations (\ref{uni1})-(\ref{uni10}) with $n=0$. A systematic study of the space of solutions of these equations will be performed elsewhere. Here we will look for the first solution one would like to identify, namely the one accounting for weak disorder. For disordered systems, the case of weak disorder is generally the best understood theoretically, because of the Harris criterion \cite{Harris} and of the possibility, in some cases, to perform perturbative calculations around the pure case. 

\begin{figure}[t]
\centering
\includegraphics[width=16cm]{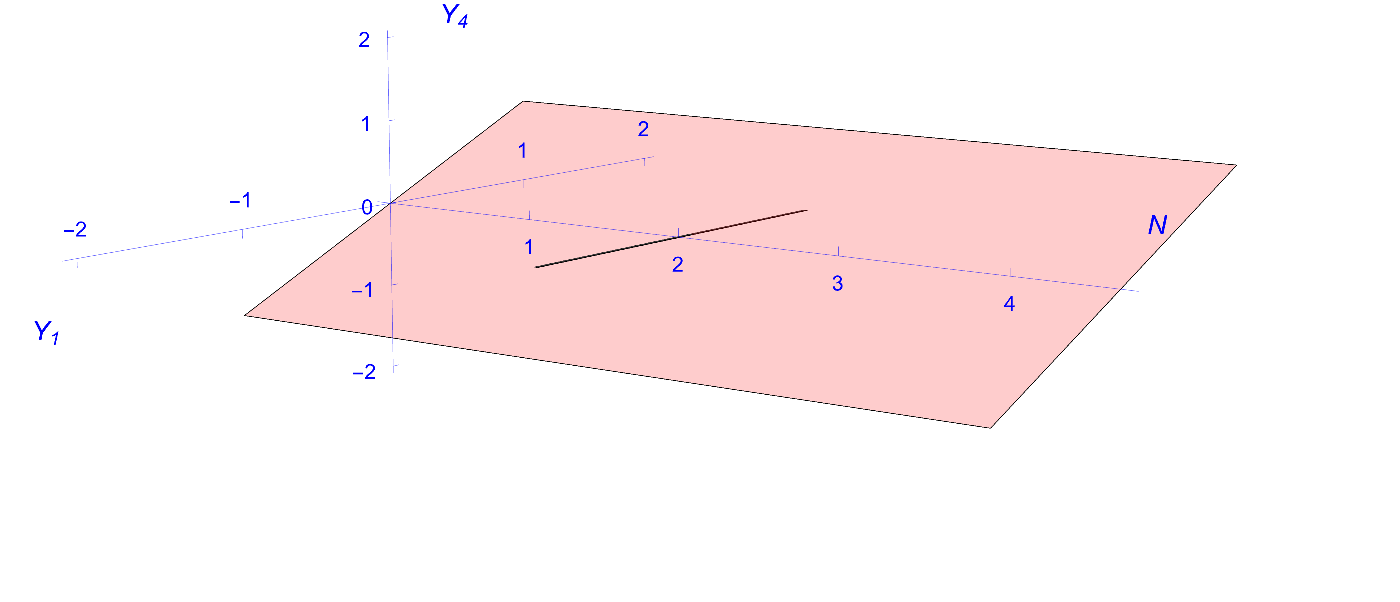}
\caption{Exact RG fixed points shown in the parameter space $Y_1$-$Y_4$, with $|Y_4|$ a measure of disorder strength. The axis $Y_4=Y_1=0$ corresponds to $N$ critical Ising ferromagnets. The surface is spanned by the line of random fixed points (\ref{random_line}) as $N$ varies. The line of fixed points at $N=2$, $Y_4=0$ is that of the pure Ashkin-Teller model.}
\label{surface}
\end{figure}

We saw that, for generic $N$, the pure system admits only the free fixed point solutions P1 of Table~\ref{pure_solutions}. Since the two-dimensional Ising pure ferromagnet corresponds to a free neutral fermion (see \cite{DfMS}), P1$_-$ with $S_0=S_2$ is the fixed point for the system of $N$ Ising pure ferromagnets. More generally, in the replicated case, the solution
\EQ
S_0=S_2=S_5=S_8=-1\,,\hspace{1.5cm}S_1=S_3=S_4=S_6=S_7=S_9=0
\label{ff}
\EN
is the fixed point for the system of $Nn$ Ising pure ferromagnets. Eqs. (\ref{uni3}), (\ref{uni6}) and (\ref{uni9}) then imply that a fixed point solution, if any, admitting a limit towards (\ref{ff}) must have $X_1=X_4=X_7=0$. For all positive values of $N$, there is a single solution of Eqs. (\ref{uni1})-(\ref{uni10}) with $n=0$ which fulfills these requirements. It reads (recall also (\ref{crossing2}))
\EQ
S_0=S_8=-1\,,\hspace{.5cm}X_1=X_4=S_7=0\,,\hspace{.5cm}Y_1=Y_4\,,\hspace{.5cm}S_2=S_5=-\sqrt{1-Y_4^2}\,,\hspace{.5cm}Y_4\in[-1,1].
\label{random_line}
\EN
Since $Y_4$ is free to vary in an interval, this solution yields a {\it line of random fixed points} for $N$ fixed (Fig.~\ref{surface}). The purely ferromagnetic solution (\ref{ff}) is recovered in the limit $Y_4\to 0$, which then is the limit of weakly disordered ferromagnetism. Conversely, the fixed points with $Y_4$ not small correspond to strong disorder.

\section{RG flows}
\label{RG_flows}
The scaling limit of the Hamiltonian (\ref{lattice}) in the purely ferromagnetic case ($J_{xy}=J>0$) is
\EQ
{\cal H}_\textrm{pure}=\sum_{a=1}^N\left[{\cal H}_a^0-\tau\int d^2x\,\varepsilon_a(x)\right]-\lambda\sum_{a\neq b}\int d^2x\,\varepsilon_a(x)\varepsilon_b(x)\,,
\label{pure}
\EN
where ${\cal H}_a^0$ and $\varepsilon_a(x)$ are the fixed point Hamiltonian (free massless neutral fermion) and the energy density field (product of the fermion components $\psi_a$ and $\bar{\psi}_a$) of the $a$-th Ising model, respectively. Random bonds $J_{xy}$ yield a random coupling $\tau(x)$. This can be handled (see \cite{Cardy_book}) passing to the replicated system with Hamiltonian
\EQ
{\cal H}_\textrm{replicas}=\sum_{a=1}^N\sum_{i=1}^n\left[{\cal H}_{a,i}^0-\int d^2x\,\tau(x)\,\varepsilon_{a,i}(x)\right]-\lambda\sum_{a\neq b}\sum_{i}\int d^2x\,\varepsilon_{a,i}(x)\varepsilon_{b,i}(x)\,,
\label{replica}
\EN
and then performing a cumulant expansion over disorder. This leads to 
\bea
{\cal H}&=&\sum_{a=1}^N\sum_{i=1}^n\left[{\cal H}_{a,i}^0-\tau\int d^2x\,\varepsilon_{a,i}(x)\right]-\int d^2x\left[\lambda_1\sum_{a}\sum_{i\neq j}\varepsilon_{a,i}(x)\varepsilon_{a,j}(x)\right.\nonumber\\
&+&\left.\lambda_2\sum_{a\neq b}\sum_{i}\varepsilon_{a,i}(x)\varepsilon_{b,i}(x)+\lambda_3\sum_{a\neq b}\sum_{i\neq j}\varepsilon_{a,i}(x)\varepsilon_{b,j}(x)\right],
\label{cumulant}
\eea
where we omitted terms involving the product of more than two energy density fields, since they are irrelevant in the RG sense. Indeed, the scaling dimension of $\varepsilon_{a,i}$ at the Ising ferromagnetic fixed point we are expanding around is $x_{\varepsilon}=1$, so that the products of two energy densities appearing in (\ref{cumulant}) are marginal fields\footnote{We also recall that in the pure two-dimensional Ising model the operator product expansion (OPE) $\varepsilon_{a,i}\cdot\varepsilon_{a,i}$ produces the identity plus irrelevant fields (see \cite{DfMS}).}. Let us call $A_\alpha(x)$ ($\alpha=1,2,3$) the sum of marginal fields conjugated to $\lambda_\alpha$ in (\ref{cumulant}). It follows from the general form of the one-loop RG equations \cite{Cardy_book} that for the couplings $\lambda_\alpha$ we have
\EQ
\frac{d\lambda_\alpha}{dt}=\sum_{\beta,\gamma}C_{\beta,\gamma}^\alpha\,\lambda_\beta\lambda_\gamma+O(\lambda^3)\,,
\label{RG}
\EN
where $t$ is the logarithmic RG scale and $C_{\beta,\gamma}^\alpha$ the coefficient of $A_\alpha$ in the OPE $A_\beta\cdot A_\gamma$. These OPE coefficients are easily determined using the combinatorial method \cite{Cardy_book}. Since $A_\alpha$ is a sum of products of two energy densities, $A_\beta\cdot A_\gamma$ produces $A_\alpha$ upon a single contraction $\varepsilon_{a,i}\cdot\varepsilon_{b,j}\to\delta_{ab}\delta_{ij}$. Performing such contractions finally yields
\bea
\frac{d\lambda_1}{dt}&=&4 (n - 2)\lambda_1^2 + 4 (N - 1) (n - 2) \lambda_3^2 + 8 (N - 1) \lambda_2 \lambda_3\,,\label{RG1}\\
\frac{d\lambda_2}{dt}&=& 4 (N - 2) \lambda_2^2 + 4 (N - 2) (n - 1) \lambda_3^2 + 8 (n - 1) \lambda_1 \lambda_3\,,\\
\frac{d\lambda_3}{dt}&=& 4 (N - 2) (n - 2) \lambda_3^2 + 8 \lambda_1 \lambda_2 + 8 (n - 2) \lambda_1 \lambda_3 + 8 (N - 2) \lambda_2 \lambda_3\,.\label{RG3}
\eea

In the pure case $\lambda_1=\lambda_3=0$ we are left with $d\lambda_2/dt=4(N-2) \lambda_2^2$, which correctly reproduces the RG equation of the Gross-Neveu model \cite{GN} ($N$ neutral fermions coupled by $O(N)$-symmetric four-fermion interaction). This equation means that $\lambda_2>0$ is marginally relevant (irrelevant) for $N>2$ ($N<2$). We know that marginality at $N=2$ persists at all orders and produces the line of fixed points of the Ashkin-Teller model\footnote{See \cite{D_AT,DG_AT} for the off-critical correlation functions in the scaling limit towards this line.}. 

Since the transition point of the pure two-dimensional Ising ferromagnet is invariant under Kramers-Wannier duality \cite{KW}, the energy density field $\varepsilon$ is odd under this duality. The transition of the pure model (\ref{pure}) is again expected to occur at a self-dual point and, since $\varepsilon_a\varepsilon_b$ is even under duality, this allows a nonzero value of $\lambda=\lambda_2$ at the transition $\tau=0$. The fact that $\lambda_2>0$ is marginally relevant for $N>2$ then implies that in this case a finite correlation length is developed and the transition is of the first order \cite{GW,GN,Fradkin,Shankar}. 

For $n=0$, (\ref{RG1})-(\ref{RG3}) are the RG equations of the disordered case. It can be checked that they reproduce those obtained in \cite{DD_AT,Dotsenko_AT} for $N=2$, in \cite{Murthy} for $N$ generic, and in \cite{Cardy_GN} for the $O(N)$-invariant case $\lambda_1=\lambda_3$. These works considered RG flows ending in the pure Ising fixed point $\lambda_1=\lambda_2=\lambda_3=0$, which for $N>2$ are sufficient to infer a softening of the first order transition of the pure model into a continuous transition with Ising exponents. The softening of first order transitions by disorder in two dimensions has been proved in \cite{AW}. 

\begin{figure}[t]
    \centering
    \begin{subfigure}[h]{0.49\textwidth}
        \includegraphics[width=\textwidth]{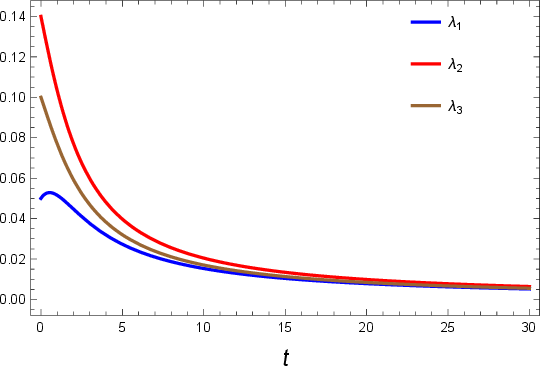}
    \end{subfigure}\hspace{.3cm}%
    \begin{subfigure}[h]{0.49\textwidth}
        \includegraphics[width=\textwidth]{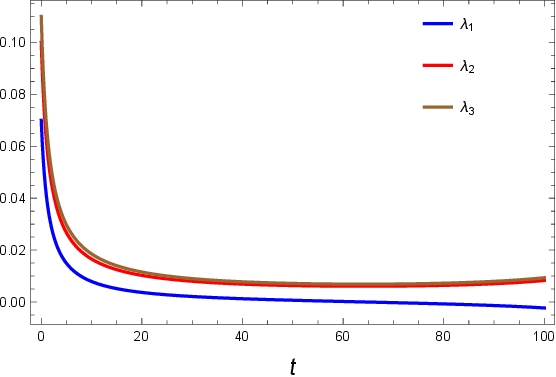}
    \end{subfigure}
    \caption{RG flows determined by Eqs.~(\ref{RG1})-(\ref{RG3}) for two different initial conditions at $N=2$ in the random case $n=0$.
    }
    \label{flows}
\end{figure}

On the other hand, the r.h.s. of (\ref{RG1})-(\ref{RG3}) with $n=0$ more generally vanish for 
\EQ
\lambda_1=0\,,\hspace{1cm}\lambda_2=\lambda_3\,,
\label{pert_line}
\EN
yielding a line of random fixed points for fixed $N$ which is not an accident of the one-loop approximation. Indeed, we know from our derivation in the scattering framework that (\ref{pert_line}) is the weak disorder limit of the exact line of fixed points (\ref{random_line}) with the identifications $Y_7\to\lambda_1$, $Y_1\to\lambda_2$ and $Y_4\to\lambda_3$. Figure~\ref{flows} shows RG flows determined by (\ref{RG1})-(\ref{RG3}) for two different initial conditions at $n=0$, $N=2$. In the left panel the couplings flow to the fixed point $\lambda_3=0$ on the line (\ref{pert_line}), namely the pure Ising fixed point. In the right panel the couplings flow towards a fixed point with $\lambda_3\neq 0$ on the line (\ref{pert_line}) and stay close to it for a substantial amount of the RG ``time" before moving away and eventually leaving the perturbative region $|\lambda_\alpha|\ll 1$ for values of $t$ not shown in the figure. Similar flows are observed for the other values of $N$. The conclusion is that the pure Ising fixed point $\lambda_3=0$ is the only one on the line (\ref{pert_line}) that can be asymptotically accessed at large distances {\it in the one-loop approximation}; inclusion of the higher order terms neglected in (\ref{RG1})-(\ref{RG3}) would make accessible also the other (random) fixed points on the line. To which point on the line the system flows depends on the initial conditions, i.e. on the system parameters, and this amounts to nonuniversality.

\section{Discussion}

The disordered model with $N>2$ was studied numerically in \cite{Katzgraber1,Katzgraber2,Vojta_3color} ($N=3$) and \cite{Vojta_4color} ($N=4$). Refs. \cite{Katzgraber1,Katzgraber2} found critical exponents varying with the system parameters, while \cite{Vojta_3color,Vojta_4color} concluded for Ising critical exponents. We now see that these two outcomes do not exclude each other and can both be realized depending on the choice of system parameters. In particular, the results of \cite{Katzgraber1,Katzgraber2} indicating nonuniversal critical behavior are no longer surprising, since we have shown exactly the existence of a line of random fixed points.  

For $N=2$, the perturbative conclusion of \cite{DD_AT,Dotsenko_AT} that the disordered system renormalizes (up to logarithms) towards the pure Ising fixed point pointed to a violation of the Harris criterion. Indeed, according to the latter, since the energy density scaling dimension $x_\varepsilon$ varies continuously along the line of fixed points of the pure Ashkin-Teller model, the same critical exponents of this pure model should be observed in the region $x_\varepsilon>1$ in which weak disorder is irrelevant ($2x_\varepsilon>2$). The conflict is solved by our result that one-loop RG flows have to be considered in light of the existence of an exact line of random fixed points, and that higher perturbative orders play a role. The region $x_\varepsilon<1$ in which weak disorder is relevant was studied numerically in \cite{WD}. The evidence of parameter-dependent critical exponents, which in absence of theoretical support they presented with caution, is now explained by the existence of a line of fixed points also in the random case.

In the random bond Ising model ($N=1$), weak disorder is marginally irrelevant and does not change the critical exponents \cite{DD_81,DD_83,Shalaev,Ludwig}. This corresponds to the fact that only the free fermion fixed point $Y_4=0$ on the line (\ref{random_line}) is physical at $N=1$ ($Y_4\neq 0$ requires $a\neq b$, i.e. $N\neq 1$). On the other hand, simulations studying spin glass properties \cite{BY} consider ``overlaps" between different copies of the same system. Theoretically, these overlaps can be formulated in the limit\footnote{In a similar way, Ising cluster connectivities are described by the Potts model with $q\to 2$ states \cite{DV1,DV2}.} $N\to 1$, in which the whole line of fixed points (\ref{random_line}) is defined and extends to strong disorder. It is then interesting to notice that there is in the literature a simulation-based debate about possible nonuniversal critical behavior in the two-dimensional Ising spin glass depending, in particular, on the disorder distribution \cite{KLC}-\cite{Weigel}.

The line of fixed points (\ref{random_line}) is likely to have implications for other realizations of two-dimensional random criticality. As a potentially interesting example, we mention that discrepancies between experimental and numerical results suggested to look for variants to the basic network model \cite{CC} used to simulate the plateau transition in the integer quantum Hall effect. For one of these variants \cite{GKNS}, numerical evidence of the presence of a line of fixed points has been presented in \cite{KNS}.

\end{document}